\RequirePackage{fix-cm}
\documentclass[twocolumn, epjc3, comma, sort&compress, natbib]{svjour3}
\bibpunct{[}{]}{,}{n}{}{,} % to get "[numbered]" references from natbib

\journalname{Eur. Phys. J. C}

\usepackage{latexsym}
\usepackage{amsmath}
\usepackage{amssymb}
\usepackage{amsfonts}
\usepackage{CJKutf8}

\usepackage[mathscr,scaled=1.15]{urwchancal}
\DeclareFontFamily{OT1}{pzc}{}
\DeclareFontShape{OT1}{pzc}{m}{it}%
{<-> s * [1.15] pzcmi7t}{}
\DeclareMathAlphabet{\mathpzc}{OT1}{pzc}{m}{it}

\usepackage{color}

\usepackage{supertabular}
\usepackage{placeins}
\usepackage{epsfig}
\usepackage{graphicx}

\definecolor{purple}{rgb}{0.5,0,0.5}
\definecolor{blue}{rgb}{0.0,0,0.9}
\definecolor{prdblue}{rgb}{0.133,0.118,0.498}
\usepackage[colorlinks=true, pdfstartview=FitV, linkcolor=prdblue, citecolor= prdblue, urlcolor=prdblue]{hyperref}

\makeatletter

\setbox0\hbox{$\xdef\scriptratio{\strip@pt\dimexpr
    \numexpr(\sf@size*65536)/\f@size sp}$}

\newcommand{\scriptveryshortarrow}[1][3pt]{{%
    \hbox{\rule[\scriptratio\dimexpr\fontdimen22\textfont2-.2pt\relax]
               {\scriptratio\dimexpr#1\relax}{\scriptratio\dimexpr.4pt\relax}}%
   \mkern-4mu\hbox{\let\f@size\sf@size\usefont{U}{lasy}{m}{n}\symbol{41}}}}

\makeatother

\begin{document}

\begin{CJK}{UTF8}{song}

%Title of paper
%\title{Symmetry-preserving calculation of the pion's valence-quark distribution}
%\title{Unification of continuum and lattice predictions of the pion's valence distribution}
%\title{Symmetry, symmetry breaking, and pion parton distributions}
\title{$\,$\\[-6ex]\hspace*{\fill}{\normalsize{\sf\emph{Preprint no}.\ NJU-INP 079/23}}\\[1ex]%
Developing predictions for pion fragmentation functions}

\author{H.-Y.~Xing\thanksref{NJU,INP}%
    $\,^{\href{https://orcid.org/0000-0002-0719-7526}{\textcolor[rgb]{0.00,1.00,0.00}{\sf ID}}}$
    \and
        Z.-Q.~Yao\thanksref{ECT}%
       $\,^{\href{https://orcid.org/0000-0002-9621-6994}{\textcolor[rgb]{0.00,1.00,0.00}{\sf ID}}}$
    \and
        B.-L.~Li\thanksref{USST}%
    $\,^{\href{https://orcid.org/0000-0002-6348-604X}{\textcolor[rgb]{0.00,1.00,0.00}{\sf ID}}}$
    \and
    D.~Binosi\thanksref{ECT}%
    $\,^{\href{https://orcid.org/0000-0003-1742-4689}{\textcolor[rgb]{0.00,1.00,0.00}{\sf ID}}}$
    \and
        Z.-F.~Cui\thanksref{NJU,INP}%
       $\,^{\href{https://orcid.org/0000-0003-3890-0242}{\textcolor[rgb]{0.00,1.00,0.00}{\sf ID}}}$
    \and
        C.~D.~Roberts\thanksref{NJU,INP}%
       $\,^{\href{https://orcid.org/0000-0002-2937-1361}{\textcolor[rgb]{0.00,1.00,0.00}{\sf ID}}}$
}

\authorrunning{H.-Y.~Xing \emph{et al}.} % if too long for running head

\institute{School of Physics, Nanjing University, Nanjing, Jiangsu 210093, China \label{NJU}
           \and
           Institute for Nonperturbative Physics, Nanjing University, Nanjing, Jiangsu 210093, China \label{INP}
           \and
           European Centre for Theoretical Studies in Nuclear Physics
            and Related Areas, \\\hspace*{1ex}Villa Tambosi, Strada delle Tabarelle 286, I-38123 Villazzano (TN), Italy \label{ECT}
           \and
           Department of Physics, University of Shanghai for Science and Technology, Shanghai 200093, China \label{USST}
\\[1ex]
Email:
\href{mailto:phycui@nju.edu.cn}{phycui@nju.edu.cn} (Z.-F. Cui);
\href{mailto:cdroberts@nju.edu.cn}{cdroberts@nju.edu.cn} (C. D. Roberts)
            }

\date{2023 Nov 01}
%\date{2023 Sep 27}

\maketitle

\end{CJK}

\begin{abstract}
Exploiting crossing symmetry, the hadron scale pion valence quark distribution function is used to predict the kindred elementary valence quark fragmentation function (FF).  This function defines the kernel of a quark jet fragmentation equation, which is solved to obtain the full pion FFs.  After evolution to a scale typical of FF fits to data, the results for quark FFs are seen to compare favourably with such fits.  However, the gluon FF is markedly different.  Notably, although FF evolution equations do not themselves guarantee momentum conservation, inclusion of a gluon FF which, for four quark flavours, distributes roughly 11\% of the total light-front momentum fraction, is sufficient to restore momentum conservation under evolution.  Overall, significant uncertainty is attached to FFs determined via fits to data; hence, the features of the predictions described herein could potentially provide useful guidance for future such studies.
\end{abstract}
%%
%%Keywords:
%%proton \sep
%%magnetic charge radius \sep
%%electric charge radius \sep
%%emergence of mass \sep
%%lepton-hadron scattering \sep
%%strong interactions in the standard model of particle physics

\maketitle

%%%%%%%%%%%%%%%%%%%%%%%%%%%%%%%%%%%%%%%%%%%%%%%%%%%%%%%%%%%%%%%%%%%%%%%%%%%%%%%%%%%%%%%%%%%%%%%%%%%%%%%%%%%%%%%%%%%%%%%
% 4500 words

\section{Introduction}
%\noindent\emph{1.$\;$Introduction}.
%
High energy interactions often produce jets of energetic hadrons, with nearly parallel longitudinal momenta and relatively small transverse momenta.  The first such jets were seen in cosmic ray events \cite{Edwards:1957}; then, on earth, using particle accelerators.  Today, these jets are normally understood to originate with gluon and quark partons that, after being produced in the initial collision, escape the interaction region and, driven by confinement forces, fragment into a shower of colourless hadrons \cite{Field:1976ve, Field:1977fa, Altarelli:1981ax, Ellis:1991qj, Metz:2016swz, Chen:2023kqw}.  These hadronisation processes are described by fragmentation functions (FFs), which may be interpreted as probability densities.   For instance, $D_u^{\pi^+}\!(z;\zeta) dz$ is the probability that, in an interaction characterised by an energy scale $\zeta$, a $u$ quark escaping the collision region produces a positively charged pion, giving up a fraction $z$ of its pre-emission light-front momentum.
This product is sometimes reinterpreted, equivalently, as the number of positive pions ``inside'' the $u$ quark within the identified momentum fraction range at the scale $\zeta$.
In ideal circumstances, FFs are independent of the parton production process; and the implicit connection with confinement means that knowledge and understanding of FFs may reveal novel aspects of emergent hadron mass (EHM) \cite{Roberts:2021nhw, Binosi:2022djx, deTeramond:2022zcm, Salme:2022eoy, Papavassiliou:2022wrb, Ding:2022ows, Ferreira:2023fva, Krein:2023azg}.

Fragmentation functions are typically extracted in global fits to selections of hadron production data \cite{Hirai:2007cx, deFlorian:2014xna, Bertone:2017tyb}; but, existing results have large uncertainties.  This is an issue because precise knowledge will be necessary if best use is to be made of data obtained at forefront and next-generation accelerator facilities \cite{Aguilar:2019teb, Anderle:2021wcy, Arrington:2021biu, Schnell:2022nlf, Accardi:2023chb}.

In such circumstances, both the need for and importance of sound theoretical predictions are magnified.  However, like parton distribution functions (DFs), FFs are nonperturbative objects, not calculable in perturbation theory; and, hitherto, no realistic results have been available.  Herein, we address this problem using continuum Schwinger function methods (CSMs) \cite{Roberts:2015lja, Eichmann:2016yit, Fischer:2018sdj, Qin:2020rad}.

We introduce a quark jet hadroproduction equation in Sec.\,\ref{SecJet}.
This is followed by a short description of pion FFs in Sec.\,\ref{SecPiFFs}, as appropriate to the hadron scale, $\zeta=\zeta_{\cal H}$, whereat, for DFs, valence degrees-of-freedom carry all the hadron's properties \cite{Yin:2023dbw}.
Crossing symmetry and the Drell-Levy-Yan (DLY) relation \cite{Drell:1969jm, Drell:1969wd, Drell:1969wb, Gribov:1972ri, Gribov:1972rt} are highlighted in Sec.\,\ref{SecDLY}.
Section~\ref{SecSCI} illustrates, largely algebraically, some of the introduced concepts using a symmetry-preserving regularisation of a contact interaction (SCI).
A scale evolution scheme for DFs and FFs is explained in Sec.\,\ref{SecFFE} and illustrated using SCI inputs.
Results for FFs obtained using realistic CSM predictions are described in Sec.\,\ref{SecFFRealistic}, wherein they are also compared with existing fits to data.
Section~\ref{Epilogue} presents a summary and perspective.

\section{Jet Equation}
\label{SecJet}
A plausible path to the calculation of FFs was outlined long ago \cite[Fig.\,1]{Field:1977fa}.  The idea is straightforward.  Beginning with knowledge of the function that describes the first fragmentation event for parton $\mathpzc p$ generating hadron $h$ with momentum fraction $z$ in the process, \emph{i.e}., the elementary FF, $d_{\mathpzc p}^h(z)$, then the complete FF, $D_{\mathpzc p}^h(z)$ -- the probability density for finding $h$ with momentum fraction $z$ in the jet, is obtained via a recursion relation that resums the entire series of such events:
\begin{equation}
\label{JetEquation}
D_{\mathpzc p}^h(z) = d_{\mathpzc p}^h(z) + \int_z^1 (dy/y)\, d_{\mathpzc p}^h(1-z/y) D_{\mathpzc p}^h(y) \,.
\end{equation}
Since there is unit probability that the parton generates a hadron with some momentum fraction, then the elementary FF is normalised, $\int_0^1 dz \, d_{\mathpzc p}^h(z)=1$.
Moreover, %on $z\simeq 1$, $D_{\mathpzc p}^h(z)\approx d_{\mathpzc p}^h(z)$
\begin{equation}
\label{largez}
D_{\mathpzc p}^h(z) \stackrel{z\simeq 1}{\approx} d_{\mathpzc p}^h(z)
\end{equation}
because if the parton gives all its momentum to $h$, then there is none left to contribute to a cascade.
%%Finally, it is straightforward to establish algebraically that the solution of Eq.\,\eqref{JetEquation} satisfies
%%\begin{equation}
%%\label{zAunity}
%%\int_0^1dz\,z\,D_{\mathpzc p}^h(z) = 1\,.
%%\end{equation}
It is readily established algebraically that the solution of Eq.\,\eqref{JetEquation} satisfies
$\int_0^1dz\,z\,D_{\mathpzc p}^h(z) = 1$.

Equation~\eqref{JetEquation} is a Volterra integral equation of the second kind, whose kernel is defined by $d_{\mathpzc p}^h(1-z)$.  Given Eq.\,\eqref{largez}, then the integral must vanish faster than $d_{\mathpzc p}^h(z)$ as $z\to 1$.  This places a constraint on the $z\simeq 0$ behaviour of admissible forms of the elementary FF.  The restriction can be exposed by noting that, with complete generality, one may write
\begin{equation}
d_{\mathpzc p}^h(z) = {\mathpzc n}(\alpha,\beta;{\mathpzc f})  \, z^\alpha (1-z)^\beta {\mathpzc f}(z)\,,
\end{equation}
where ${\mathpzc f}(z)$ is a polynomial, positive-definite and regular on $0\leq z\leq 1$, and ${\mathpzc n}(\alpha,\beta;{\mathpzc f})$ is a constant that ensures unit normalisation. %of $d_{\mathpzc p}^h(z)$.
Then, evaluating the integral on the shrinking domain obtained as $z\to 1$, one finds that
\begin{equation}
\alpha>-1
\label{alphabound}
\end{equation}
is required in order to ensure Eq.\,\eqref{largez}.

To go further, it is necessary to know the elementary FF.  Herein, we focus on $h=\pi$, \emph{viz}.\ pion FFs, for which analyses using CSMs are most straightforward \cite{Roberts:2021nhw}.  Generalisations to other systems are possible and will be provided elsewhere.
Applications of Eq.\,\eqref{JetEquation} to such problems do already exist \cite{Peng:2023mxp}.

\section{Pion fragmentation functions}
\label{SecPiFFs}
Considering the ${\mathpzc G}$-parity symmetry limit \cite{Lee:1956sw} and temporarily ignoring gluon and heavier quark degrees-of-freedom, then one can distinguish three hadron scale ($\zeta=\zeta_{\cal H}$) quark-to-pion fragmentation functions:
\begin{subequations}
\label{Favoured}
\begin{align}
D_{\mathpzc u}^{\pi^+}(z) & = D_{\bar{\mathpzc d}}^{\pi^+}(z)
= D_{\bar{\mathpzc u}}^{\pi^-}(z) = D_{{\mathpzc d}}^{\pi^-}(z) \,,\\
D_{\bar{\mathpzc u}}^{\pi^+}(z) & = D_{{\mathpzc d}}^{\pi^+}(z) =
D_{{\mathpzc u}}^{\pi^-}(z) = D_{\bar {\mathpzc d}}^{\pi^-}(z)\,, \\
D_{{\mathpzc u}}^{\pi^0}(z) & = D_{{\mathpzc d}}^{\pi^0}(z)
= D_{\bar{\mathpzc u}}^{\pi^0}(z)  = D_{\bar{\mathpzc d}}^{\pi^0}(z)\,.
\end{align}
\end{subequations}
%%  -- respectively, favoured, unfavoured, neutral
The first row describes the cases in which the hadronising quark or antiquark can be a valence degree-of-freedom in the produced pion (favoured);
the second row, those situations when it cannot (unfavoured);
and the third row, when any initial quark or antiquark fla\-vour can be a valence part of the emitted pion (neutral).
These functions are not all independent, as is made clear, \emph{e.g}., by considering the initial (favoured) step in the $\pi^+$ chain, \emph{viz}.\ $u \to  d \pi^+$: plainly, the next step in $\pi^+$ production must involve unfavoured fragmentation: $d \to u \pi^+$.

Again exploiting ${\mathpzc G}$-parity symmetry, the pion FFs can be decoupled by introducing $q$-singlet and $q$-non\-sing\-let FFs, respectively:
\begin{subequations}
\begin{align}
D_{S_q}^\pi(z) & = \frac{3}{2} \left[ D_{\mathpzc q}^{\pi^+}(z) + D_{\bar{\mathpzc q}}^{\pi^+}(z)\right]\,. \\
D_{N_q}^\pi(z) & = \frac{3}{2} \left[ D_{\mathpzc q}^{\pi^+}(z) - D_{\bar{\mathpzc q}}^{\pi^+}(z)\right] \,.
\end{align}
\end{subequations}
(The $3/2$ is just an isospin Clebsch-Gordon factor.)
These functions can be obtained by solving \cite{Ito:2009zc}
\begin{subequations}
\label{SNSFFEqs}
\begin{align}
D_{S_q}^\pi(z) & =  d_\pi(z) + \int_z^1 dy\, d_\pi(1-z/y)\frac{1}{y} D_{S_q}^\pi(y)\,, \label{EqSFF}
 \\
D_{N_q}^\pi(z) & = d_\pi(z) - \frac{1}{3} \int_z^1 dy\, d_\pi(1-z/y)\frac{1}{y} D_{N_q}^\pi(y)\,. \label{EqNSFF}
\end{align}
\end{subequations}
As noted above, $\int_0^1 dz\,z D_{S_q}^\pi(z) = 1$; furthermore, \linebreak $\int_0^1 dz\, D_{N_q}^\pi(z) = 3/4$.
In terms of the solutions:
\begin{subequations}\label{FFsolutions}
\begin{align}
D_{\mathpzc u}^{\pi^+}(z;\zeta_{\cal H}) & = \frac{1}{3} \left[ D_{S_{\mathpzc u}}^\pi(z) + D_{N_{\mathpzc u}}^\pi(z) \right]\,, \\
D_{\bar{\mathpzc u}}^{\pi^+}(z;\zeta_{\cal H}) & =  \frac{1}{3} \left[ D_{S_{\mathpzc u}}^\pi(z) - D_{N_{\mathpzc u}}^\pi(z) \right]\,, \\
D_{{\mathpzc u}}^{\pi^0}(z;\zeta_{\cal H})  & = \frac{1}{3} D_{S_{\mathpzc u}}^\pi(z)\,.
\end{align}
\end{subequations}

\section {Drell-Levy-Yan Relation}
\label{SecDLY}
Crossing symmetry in quantum field theory means that elementary fragmentation functions may be viewed as a timelike analogue of parton DFs \cite{Drell:1969jm, Drell:1969wd, Drell:1969wb, Gribov:1972ri, Gribov:1972rt}.  Practically, this translates into the following correspondence -- the DLY relation:
\begin{equation}
\label{DLYR}
d_{\mathpzc p}^\pi(z;\zeta) \propto z {\mathpzc p}^\pi(1/z;\zeta)\,,
\end{equation}
where ${\mathpzc p}^h(x;\zeta)$ is the ${\mathpzc p}$-parton distribution function in $h$ at resolving scale $\zeta$.
The DLY relation has been exploited in the some of the few available model calculations of FFs \cite{Ito:2009zc, Nam:2012af}.

It is worth noting that Eq.\,\eqref{DLYR} entails that all manifestations of EHM in ${\mathpzc p}^h$ are also expressed in the source function which drives fragmentation of ${\mathpzc p}$ into $h$.  This information flows into the full fragmentation function via Eq.\,\eqref{SNSFFEqs}.  Thus, for instance and not unexpectedly, perhaps, the seeds of confinement, as expressed in hadronisation, are already to be found in the wave functions of the hadrons involved.

Owing to scaling violations in quantum chromodynamics, one must identify the scale at which Eq.\,\eqref{DLYR} is valid.  The natural value is $\zeta = \zeta_{\cal H}$, whereat dressed valence degrees-of-freedom carry all properties of the pion \cite{Yin:2023dbw}: in-pion sea and glue DFs vanish at $\zeta_{\cal H}$.
This choice is logical for, \emph{inter alia}, the following reasons.
A realistic pion valence quark DF, ${\mathpzc q}^\pi(x;\zeta_{\cal H})$, is symmetric under $x\leftrightarrow (1-x)$; vanishes at the endpoints, $x=0,1$; and can be written in a form that ensures Eq.\,\eqref{DLYR} produces an elementary FF which vanishes at $z=0,1$.  These things guarantee that Eq.\,\eqref{alphabound} is satisfied without tuning; hence, Eq.\,\eqref{JetEquation} possesses a robust solution.  QCD scaling violations mean that such cannot be guaranteed for $\zeta>\zeta_{\cal H}$.

Returning now to Eq.\,\eqref{DLYR}, it can be seen to entail that the $z \simeq 1$ behaviour of the $\zeta=\zeta_{\cal H}$ elementary FF matches that of the related valence quark DF on $x\simeq 1$.  In QCD, this means \cite{Brodsky:1994kg, Yuan:2003fs, Holt:2010vj, Cui:2021mom, Cui:2022bxn, Lu:2022cjx}: $d_{\mathpzc p}^\pi(z\simeq 1;\zeta_{\cal H}) \propto (1-z)^2$.  It furthermore follows from Eq.\,\eqref{largez} that the same is true of $D^\pi(z;\zeta_{\cal H})$.  Since the large-$z$ power increases under evolution, then any QCD-consistent favoured FF should express the following behaviour:
\begin{equation}
\label{FFzone}
D_{{S_q},{N_q}}^\pi(z;\zeta) \stackrel{z\simeq 1}{\propto} (1-z)^{2+\gamma(\zeta)},
\end{equation}
where $\gamma(\zeta>\zeta_{\cal H}) \geq 0$ grows logarithmically with $\zeta$.  The powers on glue and sea FFs are, respectively, one and two units greater \cite{Brodsky:1994kg, Yuan:2003fs, Holt:2010vj, Cui:2021mom, Cui:2022bxn, Lu:2022cjx}.  However, as with analyses of data that attempt to infer DFs, these constraints are typically overlooked in phenomenological extractions of FFs.

\section{Contact Interaction}
\label{SecSCI}
It is useful to illustrate some of the preceding remarks using the symmetry preserving regularisation of a vector\,$\times$\,vector contact interaction (SCI) \cite{GutierrezGuerrero:2010md}.  In the chiral limit, \emph{i.e}., when the quark current masses are zero, it yields the following hadron scale valence quark DF \cite{Zhang:2020ecj}:
%%\begin{equation}
${\mathpzc u}^\pi(x;\zeta_{\cal H}) = 1$;
%% \end{equation}
and, via Eq.\,\eqref{DLYR}:
\begin{equation}
\label{SCIEFF}
d_{\mathpzc u}^\pi(z;\zeta_{\cal H}) = 2 z\,.
\end{equation}

\begin{figure}[t]
\vspace*{0ex}

\leftline{\hspace*{0.5em}{{\textsf{A}}}}
\vspace*{-2ex}
\centerline{\includegraphics[width=0.85\columnwidth]{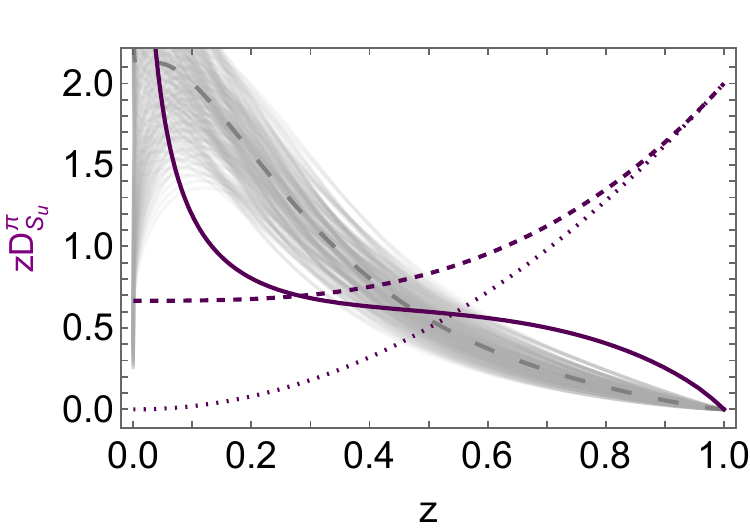}}
\vspace*{2ex}

\leftline{\hspace*{0.5em}{{\textsf{B}}}}
\vspace*{-2ex}
\centerline{\includegraphics[width=0.85\columnwidth]{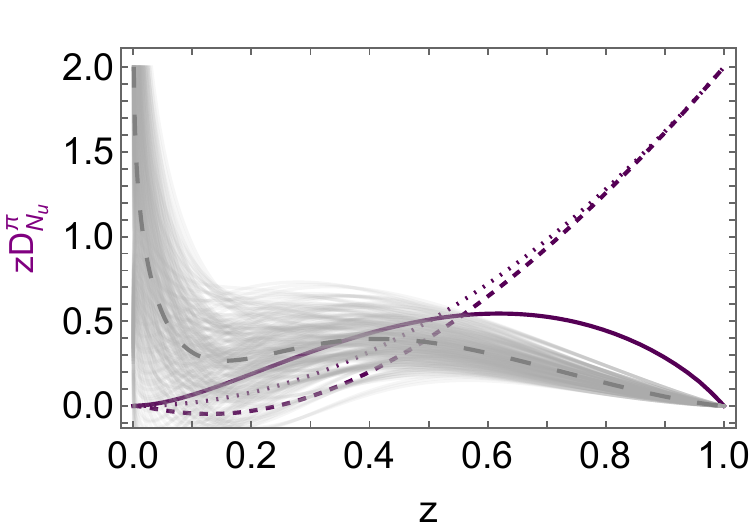}}

%\begin{tabular}{c}
%\includegraphics[width=0.85\columnwidth]{F2.pdf}
%\end{tabular}
%\vspace*{-0.25cm}
\caption{\label{FSCI}
\emph{Panel A}.
Contact interaction singlet pion fragmentation function.
$z D_S^\pi(z;\zeta_{\cal H})$ -- from Eq.\,\eqref{SolSFF} [dashed purple curve], obtained by solving Eq.\,\eqref{EqSFF} with kernel defined by Eq.\,\eqref{SCIEFF} [dotted purple curve].
$z D_S^\pi(z;\zeta_1=1\,{\rm GeV})$ [solid purple curve], \emph{i.e}., after evolution, as described in Sec.\,\ref{SecFFE}.  For comparison, the long-dashed grey curve within like-coloured band shows
$[D_{\mathpzc u}^{\pi^+}+D_{\bar {\mathpzc u}}^{\pi^+}]$
constructed using the fits (LO) in Ref.\,\cite[HKNS]{Hirai:2007cx}.
\emph{Panel B}. Analogous curves for the nonsinglet pion fragmentation function: $z D_N^\pi(z)$.% -- from Eq.\,\eqref{SolNSFF} [solid purple curve], obtained by solving Eq.\,\eqref{EqNSFF} with kernel defined by Eq.\,\eqref{SCIEFF} [dotted purple curve].
}
\end{figure}

Inserting Eq.\,\eqref{SCIEFF} into Eqs.\,\eqref{SNSFFEqs}, one finds
\begin{subequations}\label{SolnFFs}
\begin{align}
D_{S_{\mathpzc u}}^{\pi}(z) & = \frac{4 z^2}{3} + \frac{2}{3z}\,, \label{SolSFF}\\
D_{N_{\mathpzc u}}^{\pi}(z) & = \frac{2}{5} \sqrt{z} \left[\sqrt{15} \sin \left(\sqrt{\frac{5}{12}} \log (z)\right) \right. \nonumber \\
   & \qquad \left. +5 \cos
 \left( \sqrt{\frac{5}{12}} \log (z)\right)\right]\,. \label{SolNSFF}
\end{align}
\end{subequations}
These solutions are drawn in Fig.\,\ref{FSCI}.
Evidently, as expected -- Eq.\,\eqref{largez}, both singlet and nonsinglet FFs approach the elementary FF on $z\simeq 1$.
It should nevertheless be observed that the singlet FF exhibits a nonintegrable singularity on $z\simeq 0$.
Notably, the solution of the singlet equation is everywhere enhanced above the elementary FF, whereas the nonsinglet solution is suppressed.
Each of these features is robust, \emph{i.e}., independent of details about the form of $d_{\mathpzc p}^\pi(z;\zeta_{\cal H})$.

\section{Fragmentation function evolution}
\label{SecFFE}
In considering the expression of QCD scaling violations in FFs, it is worth recalling the evolution equations for DFs. Herein, we adopt the all-orders scheme explained in Ref.\,\cite{Yin:2023dbw}, which has proved efficacious in numerous applications, \emph{e.g}., delivering unified predictions for all pion, kaon, and proton (unpolarised and polarised) DFs that agree with much available data \cite{Cui:2020tdf, Lu:2022cjx, Cheng:2023kmt}.  In this case, illustrating by supposing $n_f$ flavours of massless quarks, the scale evolution of singlet, $\Sigma_\pi(x;\zeta)$, and glue, ${\mathpzc g}_\pi(x,\zeta)$, DFs in the pion is described by the following integrodifferential equations, written for $\breve\Sigma_\pi(x;t)= x\Sigma_\pi(x;t)$, $\breve{\mathpzc g}_\pi(x;t)=x {\mathpzc g}_\pi(x;t)$:
{\allowdisplaybreaks
\begin{subequations}
\label{DFEvolution}
\begin{align}
    \frac{d}{dt} \breve\Sigma_\pi(x;t) & = \frac{\alpha(t)}{2\pi}
    \int_x^1\,dz
    \left[ P_{qq}(z) \breve\Sigma_\pi(x/z;t) \right. \nonumber \\& \qquad \left. + 2 n_f P_{qg}(z) \breve{\mathpzc g}_\pi(x/z;t)    \right]\,,\\
\frac{d}{dt} \breve{\mathpzc g}_\pi(x;t) & = \frac{\alpha(t)}{2\pi}
\int_x^1\,dz
    \left[ P_{gq}(z) \breve\Sigma_\pi(x/z;t) \right. \nonumber \\
    & \qquad \left. + P_{gg}(z) \breve{\mathpzc g}(x/z;t)    \right]\,,
\end{align}
\end{subequations}
where $t=\ln \zeta^2$, with $\zeta$ the scale at which the DFs are evaluated
%
%$\bar{\mathpzc S}(x;t) = x{\mathpzc S}(x;t)$, $\bar{\mathpzc G}(x;t) = x{\mathpzc G}(x;t)$ are the light-front momentum fraction weighted quark singlet and gluon DFs;
%
and the splitting functions are, with $C_F=4/3$, $T_R = 1/2$, $C_A=3$,
$\beta_0=(11 C_A - 2 n_f)/3$,
\begin{subequations}
\begin{align}
    P_{qq}(z) & = C_F\left[\frac{1+z^2}{(1-z)_+}+\frac{3}{2}\delta(1-z)\right]\,,\\
    P_{qg}(z) & = T_R [ z^2+(1-z)^2 ] \,,\\
    P_{gq}(z) & = C_F \frac{1+(1-z)^2}{z}\,,\\
    P_{gg}(z) & = 2 C_A\left[\frac{z}{(1-z)_+}+\frac{1-z}{z}+z(1-z) \right] \nonumber \\
    & \qquad +\tfrac{1}{2}\beta_0\delta(1-z)\,.
\end{align}
\end{subequations}
Here,
\begin{equation}
\int_0^1 dz \frac{f(z)}{(1-z)_+} :=
%\int_0^1 dz \frac{f(z)-f(1)}{(1-z)} \,.
\int_0^1 dz \ln (1-z) f^\prime(z) \,.
\end{equation}}

The solutions of Eqs.\,\eqref{DFEvolution} are completely determined by the pion DFs at $\zeta_{\cal H}$; and following Ref.\,\cite{Yin:2023dbw}, these functions are
$\Sigma_\pi(x;\zeta_{\cal H}) = 2 {\mathpzc V}_\pi(x;\zeta_{\cal H})$,
${\mathpzc g}(x;\zeta_{\cal H}) \equiv 0$, where ${\mathpzc V}_\pi(x;\zeta_{\cal H})$ is the hadron scale $u=d$ valence quark DF.  On $\zeta>\zeta_{\cal H}$, one obtains the valence DF by solving, with $\breve{\mathpzc V}_\pi(x;t)=x {\mathpzc V}_\pi(x;t)$,
\begin{equation}
 \frac{d}{dt} \breve{\mathpzc V}_\pi(x;t) = \frac{\alpha(t)}{2\pi}
    \int_x^1\,dz
     P_{qq}(z) \breve{\mathpzc V}(x/z;t)\,. \label{ValenceEvolution}
\end{equation}

Momentum conservation under DF evolution is guaranteed by the following identities:
{\allowdisplaybreaks
\begin{subequations}
\label{MomCons}
    \begin{align}
        0 & = \int_0^1 dz\,z [ P_{qq}(z) + P_{gq}(z)] \,, \label{MomCons1}\\
        0 & = \int_0^1 dz\, z [ 2n_f P_{qg}(z) + P_{gg}(z)] \label{MomCons2}\,.
    \end{align}
\end{subequations}}

Adapting the all-orders scheme \cite{Yin:2023dbw}, then fragmentation functions evolve according to Eqs.\,\eqref{DFEvolution}, \eqref{ValenceEvolution} with only minor changes.  Namely, for the singlet equations, the off-diagonal elements of the matrix of splitting functions are interchanged:
\begin{equation}
\label{FFsplitting}
2 n_f P_{qg}(z)  \to 2 n_f P_{gq}(z)\,, \quad
    P_{gq}(z) \to P_{qg}(z)\,;
\end{equation}
%%\begin{subequations}
%%\begin{align}
%%    2 n_f P_{qg}(z) & \to 2 n_f P_{gq}(z)\,,\\
%%    P_{gq}(z) & \to P_{qg}(z)\,;
%%\end{align}
%%\end{subequations}
so, with $D_g^\pi(x;t)$ being the pion's gluon FF, which is unfavoured in the sense of Eqs.\,\eqref{Favoured}, and $D_{S}^\pi=\sum_{q} D_{S_{\mathpzc q}}^\pi$, then, written in terms of $\breve D_{S,g}^\pi(z;t) = z \breve D_{S,g}^\pi(z;t)$,
\begin{subequations}
\label{FFEvolution}
\begin{align}
    \frac{d}{dt} \breve D_S^\pi(z;t) & = \frac{\alpha(t)}{2\pi}
    \int_z^1\,dy
    \left[ P_{qq}(y) \breve D_S^\pi(z/y;t) \right. \nonumber \\& \qquad \left. + 2 n_f P_{gq}(z) \breve D_g^\pi(z/y;t)    \right]\,,\\
\frac{d}{dt} \breve D_g^\pi(z;t) & = \frac{\alpha(t)}{2\pi}
\int_z^1\,dy
    \left[ P_{qg}(y) \breve D_S^\pi(z/y;t) \right. \nonumber \\
    & \qquad \left. + P_{gg}(y) \breve D_g^\pi(z/y;t)    \right]\,.
\end{align}
\end{subequations}

The evolution equation for $D_{N_{\mathpzc q}}^\pi(x;\zeta)$ is identical to Eq.\,\eqref{ValenceEvolution}.  Since $\int_0^1 dz\, P_{qq}(z)=0$, then the zeroth moment of $D_{N_{\mathpzc q}}^\pi(x;\zeta)$ is conserved; namely,
\begin{equation}
\int_0^1 dz\, D_{N_{\mathpzc q}}^\pi(x;\zeta) \stackrel{\forall \zeta >\zeta_{\cal H}}{=} \int_0^1 dz\, D_{N_{\mathpzc q}}^\pi(x;\zeta_{\cal H}) \,.
\end{equation}
For the pion, this amounts to a statement of isospin conservation during hadronisation.

Momentum conservation is a different matter.  Using Eqs.\,\eqref{FFsplitting} to map Eqs.\,\eqref{MomCons}, one finds
%the following analogues:
\begin{align}
 \int_0^1 dz\,z [ P_{qq}(z) + P_{qg}(z)] & = -\frac{29}{18} \,,\\
 \int_0^1 dz\,z [ 2 n_f P_{gq}(z) + P_{gg}(z)] & = \frac{29}{9} n_f\,.
\end{align}
Hence, the FF evolution equations do not conserve momentum: \emph{a priori}, the singlet FF loses momentum and the gluon FF acquires it.  In general, the loss and gain are not balanced.

There is no contradiction here.  Instead, one should appreciate that in transforming from DF to FF, one does not completely interchange the initial and final states.  With DFs, all partons involved in evolution were part of the initial state and remain so throughout the process.  Regarding FFs, on the other hand, the splitting takes one initial parton to that which enters the produced hadron and potentially infinitely many others which do not, carrying momentum away with them.  This is illustrated, \emph{e.g}., in Ref.\,\cite[Fig.\,19]{Altarelli:1981ax}.

When extracting FFs through fits to data, momentum conservation is typically enforced by requiring that the input FFs for each parton produce a collection of first Mellin moments whose sum is unity after all final state hadrons are included -- see, e.g., Ref.\,\cite[Eq.\,(11)]{Hirai:2007cx}.  Empirically, fragmentation to pion contributions are dominant but not exhaustive.

It is also important to observe that if, for instance, one begins with $D_g^\pi(x;\zeta_{\cal H})\equiv 0$, then evolution removes momentum from $D_S^\pi(x;\zeta_{\cal H})$, feeding it into $D_g^\pi(x;\zeta_{\cal H})$.  Overall, however, momentum is lost to the unresolved parton shower.

Alternatively, if one instead assumes  $D_g^\pi(z;\zeta_{\cal H})\neq 0$, then there is always a value of $\langle z\rangle_{D_g^\pi}^{\zeta_{\cal H}} = \int_0^1 dz z D_g^\pi(z;\zeta_{\cal H})$ such that
\begin{align}
\langle z\rangle_{D_S^\pi}^{\zeta}+\langle z\rangle_{D_g^\pi}^{\zeta}
\stackrel{\forall \zeta>\zeta_{\cal H}} = \langle z\rangle_{D_S^\pi}^{\zeta_{\cal H}}+\langle z\rangle_{D_g^\pi}^{\zeta_{\cal H}}\,.
%%\\
%
%%& \int_0^1 dz\, z\, [ D_S^\pi(z;\zeta)+D_g^\pi(z;\zeta) ]  \nonumber \\
%%& \stackrel{\forall \zeta>\zeta_{\cal H}} =
%%\int_0^1 dz\, z\, [ D_S^\pi(z;\zeta_{\cal H})+D_g^\pi(z;\zeta_{\cal H}) ] \,.
\label{MomBalance}
\end{align}
This is exemplified by the pion FFs obtained in Ref.\,\cite[HKNS]{Hirai:2007cx} via a fit to charged-hadron production data measured in electron + positron annihilation.  (Results from more recent fits, \emph{e.g}., Ref.\,\cite{Bertone:2017tyb}, are not materially different; so, we employ HKNS fits for comparisons herein because their parametrisations are easy to use.  They are inconsistent with Eq.\,\eqref{FFzone} and its corollaries, however.)

Working with $n_f$ flavours of massless (evolution-active) quarks, the critical value of the momentum fraction distributed by the gluon FF is
\begin{equation}
\label{GlueMomFraction}
\langle z\rangle_{D_g^\pi}^{\zeta_{\cal H}} = 1/[1+2 n_f]\,,
\end{equation}
a result that is readily established by using the first-moment evolution equation derived from Eq.\,\eqref{FFEvolution}.

Following this discussion, it will be apparent that Eq.\,\eqref{EqSFF} is incomplete.  Any singlet jet equation in QCD should properly involve gluon contributions to the cascade, because of $g \leftrightarrow q + \bar q$ mixing, and also, therefore, heavier quark + antiquark pairs, albeit to a lesser extent.   We implement this phenomenologically by writing\\[-3ex]
\begin{subequations}
\label{GlueAnsatz}
\begin{align}
D_{S_{\mathpzc q}}^\pi(z;\zeta_{\cal H}) &
\to \tilde D_{S_{\mathpzc q}}^\pi(z;\zeta_{\cal H})
+ \tilde D_{g_{\mathpzc q}}^\pi(z;\zeta_{\cal H})\,, \\
& = (1-\delta) D_{S_{\mathpzc q}}^\pi(z;\zeta_{\cal H})
+ \delta D_{g_{\mathpzc q}}^\pi(z;\zeta_{\cal H})\,,
\end{align}
\end{subequations}
with the constant $\delta\in (0,1)$ chosen to ensure Eq.\,\eqref{MomBalance} -- it should lie near the value in Eq.\,\eqref{GlueMomFraction}, and
\begin{equation}
D_{g_{\mathpzc q}}^\pi(z;\zeta_{\cal H})=D_{g_{\mathpzc u}}^\pi(z;\zeta_{\cal H}) \propto [D_{S_{\mathpzc u}}^\pi(z;\zeta_{\cal H})-D_{N_{\mathpzc u}}^\pi(z;\zeta_{\cal H})]\,,
\end{equation}
\emph{viz}.\ possessing the pointwise behaviour of the unfavoured quark FF and normalised to ensure the first Mellin moment of the right-hand side is unity.  $D_{g_{\mathpzc q}}^\pi$ is the same for each quark flavour.
We subsequently exemplify this procedure using the SCI FFs in Sec.\,\ref{SecSCI}.
Of course, contact interaction FF results and QCD (gluon-exchange interaction) evolution are not mutually consistent.  Nonetheless, the illustration provides useful insights into the impacts of evolution on hadron scale inputs.

Before proceeding, it is appropriate to note that considering $D_g^\pi(z;\zeta_{\cal H})\neq 0$ does not represent a departure from the standard all-orders evolution principle that $\zeta_{\cal H}$ is the scale whereat all properties of a given hadron are carried by its valence (quasiparticle) degrees-of-freedom \cite{Yin:2023dbw}.
This is seen by noting that, following a given collision, the fragmentation process embeds one of the produced quasiparticle partons into a particular hadron; but, irrespective of the scale, not all the collision debris can correspond to a valence degree-of-freedom in the detected hadron.
As will subsequently be seen, whilst supposing $D_g^\pi(z;\zeta_{\cal H})\neq 0$ has the potential to introduce some ambiguity into FF predictions, that is largely eliminated by enforcing Eq.\,\eqref{MomBalance} via Eq.\,\eqref{GlueAnsatz}.

Beginning with the SCI FFs described in Sec.\,\ref{SecSCI};
implementing the all-orders evolution scheme as detailed in Ref.\,\cite[Sec.\,2]{Lu:2022cjx},
including the $s$ and $c$ quark mass thresholds therein;
and setting $\delta=0.113$; then one obtains the forms of $D_{S_{\mathpzc u},N_{\mathpzc u}}(z;\zeta_1=1\,{\rm GeV})$ drawn in Fig.\,\ref{FSCI} (solid purple curves).
(The slight deviation from Eq.\,\eqref{GlueMomFraction} owes to the $s$ and $c$ quark thresholds.)
Notably, in this implementation of all-orders evolution,
\begin{equation}
\zeta_{\cal H}=0.331(2)\,{\rm GeV}
\end{equation}
is a prediction deriving from the behaviour of the process-independent charge that is explained and calculated in Ref.\,\cite{Cui:2019dwv}.  (The character, uses, and implications of QCD effective charges are reviewed elsewhere \cite{Deur:2023dzc}.)

\begin{figure}[t]
\centerline{\includegraphics[width=0.85\columnwidth]{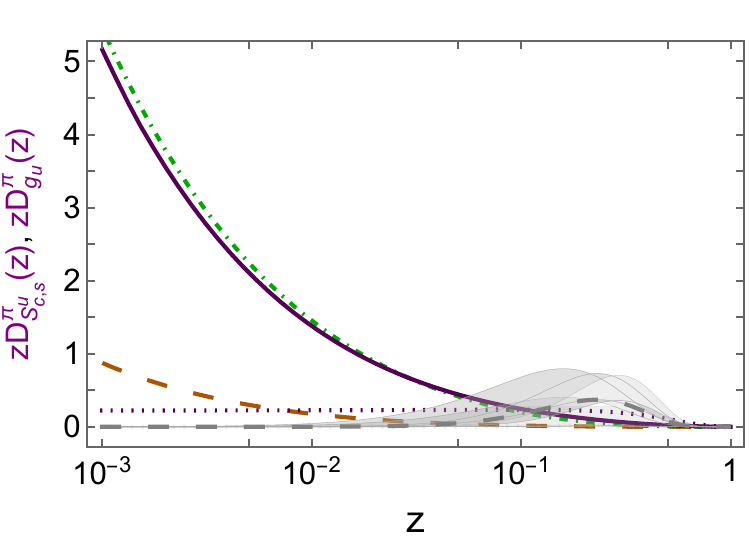}}

\caption{\label{FiggscFF}
Contact interaction.
Dotted curve -- input $D_{g_{\mathpzc u}}^\pi(z;\zeta_{\cal H})$.
After evolution $\zeta_{\cal H} \to \zeta_1 = 1\,{\rm GeV}$,
$D_{g_{\mathpzc u}}^\pi(z;\zeta_1)$ -- solid purple curve;
$D_{{S_{\mathpzc s}^{\mathpzc u}}}^\pi(z;\zeta_1)$ -- dot-dashed green;
$D_{{S_{\mathpzc c}^{\mathpzc u}}}^\pi(z;\zeta_1)$ -- long-dashed orange.
For comparison, the long-dashed grey curve within like-coloured band shows $D_{g_{\mathpzc u}}^{\pi^+}(z;\zeta_1)$ as obtained from the fit (LO) in Ref.\,\cite{Hirai:2007cx}.
}
\end{figure}

The input and evolved gluon FFs are drawn in Fig.\,\ref{FiggscFF} along with the evolution-generated $s$ and $c$ quark singlet FFs.
Using these functions, one finds
{\allowdisplaybreaks
\begin{subequations}
\label{MomBeforeAfter}
\begin{align}
\langle z \rangle_{D_{S_{\mathpzc u}}^\pi}^{\zeta_{\cal H}} & = 0.89\,,\;
\langle z \rangle_{D_{g_{\mathpzc u}}^\pi}^{\zeta_{\cal H}} = 0.11\,, \\
%\int_0^1 dz \, z  D_{S_{\mathpzc u}}^\pi(z;\zeta_{\cal H})
%1 & \int_0^1 dz\,z [ D_{g_{\mathpzc u}}^\pi(z;\zeta_{\cal H})]
\langle z \rangle_{D_{S_{\mathpzc u}}^\pi}^{\zeta_{1}} & = 0.78\,,\;
\langle z \rangle_{D_{g_{\mathpzc u}}^\pi}^{\zeta_{1}} = 0.11\,, \\
\langle z \rangle_{D_{S_{\mathpzc s}^{\mathpzc u}}^\pi}^{\zeta_{1}} & = 0.095\,,\;
\langle z \rangle_{D_{S_{\mathpzc c}^{\mathpzc u}}^\pi}^{\zeta_{1}} = 0.011\,,
\end{align}
\end{subequations}
where the terms in the last row count, respectively, the momentum fractions in $s+\bar s$ and $c+\bar c$ deriving from the fragmenting $u$ quark.
}

The grey bands in Figs.\,\ref{FSCI}, \ref{FiggscFF} represent the fits (LO) in Ref.\,\cite[HKNS]{Hirai:2007cx}.  It was observed therein that the gluon FF, in particular, is very difficult to constrain using available data.  Notwithstanding the manifest pointwise differences between evolved SCI results and the HKNS fits, the latter yield
$\langle z \rangle_{D_{{\mathpzc u}}^{\pi^+}+D_{\bar{\mathpzc u}}^{\pi^+}}^{\zeta_{1}} = 0.80(9)$,\linebreak
$\langle z \rangle_{D_{g_{\mathpzc u}}^{\pi^+}}^{\zeta_{1}} = 0.115(111)$,
values commensurate with the SCI results.  This indicates that, at least at present, the entirely algebraic SCI is a fair starting point for developing intuition about FFs.

\section{Fragmentation functions from realistic distribution functions}
\label{SecFFRealistic}
\begin{figure}[t]
\vspace*{0ex}

\leftline{\hspace*{0.5em}{{\textsf{A}}}}
\vspace*{-2ex}
\centerline{\includegraphics[width=0.85\columnwidth]{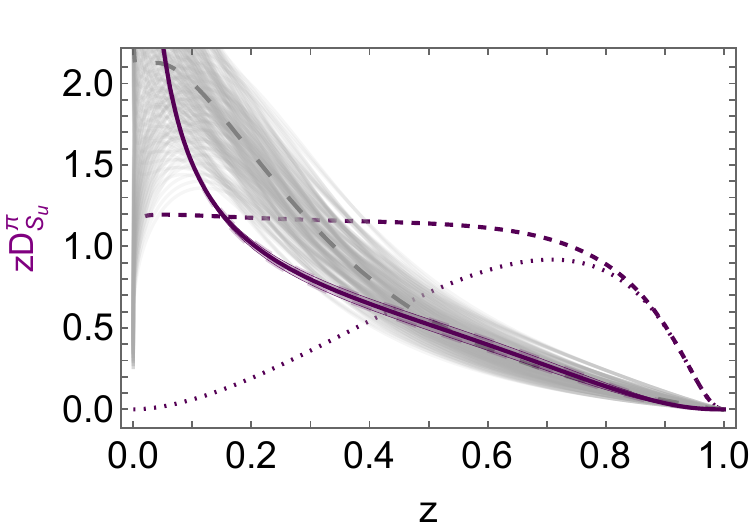}}
\vspace*{2ex}

\leftline{\hspace*{0.5em}{{\textsf{B}}}}
\vspace*{-2ex}
\centerline{\includegraphics[width=0.85\columnwidth]{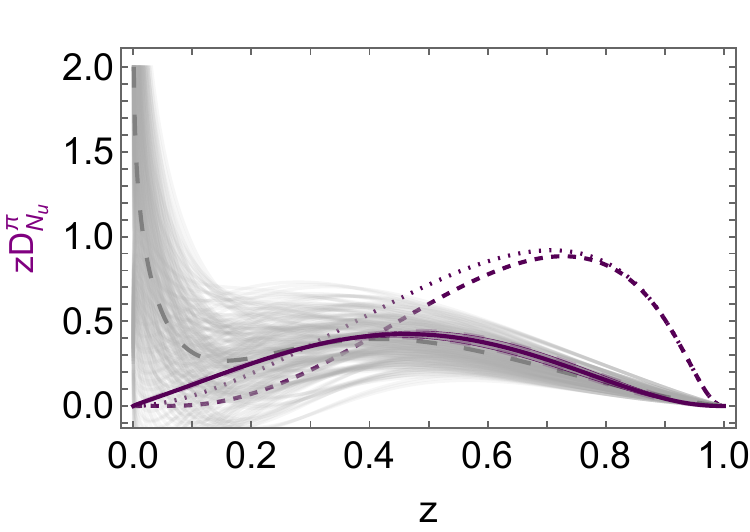}}

%\begin{tabular}{c}
%\includegraphics[width=0.85\columnwidth]{F2.pdf}
%\end{tabular}
%\vspace*{-0.25cm}
\caption{\label{FReal}
\emph{Panel A}.
CSM prediction for the singlet pion fragmentation function.
$z D_S^\pi(z;\zeta_{\cal H})$ -- discussed in connection with Eq.\,\eqref{EqJetSolutionsA} [dashed purple curve], obtained by solving Eq.\,\eqref{EqSFF} with kernel defined by Eq.\,\eqref{EqElemFF} [dotted purple curve].
$z D_S^\pi(z;\zeta_1)$ [solid purple curve], \emph{i.e}., after evolution, as described in Sec.\,\ref{SecFFE}.  The bracketing purple band indicates the response to $\zeta_{\cal H} \to (1\pm 0.05)\zeta_{\cal H}$.
For comparison, the long-dashed grey curve within like-coloured band shows
$[D_{\mathpzc u}^{\pi^+}+D_{\bar {\mathpzc u}}^{\pi^+}]$
constructed using the fits (LO) in Ref.\,\cite[HKNS]{Hirai:2007cx}.
\emph{Panel B}. Analogous curves for the nonsinglet pion fragmentation function: $z D_N^\pi(z)$.% -- from Eq.\,\eqref{SolNSFF} [solid purple curve], obtained by solving Eq.\,\eqref{EqNSFF} with kernel defined by Eq.\,\eqref{SCIEFF} [dotted purple curve].
}
\end{figure}

In Ref.\,\cite{Cui:2020tdf}, CSMs were used to calculate the hadron-scale valence quark DF, with a result that can reliably be expressed in the following algebraic form:
\begin{equation}
\label{EqValenceDF}
{\mathpzc u}^\pi(x;\zeta_{\cal H}) =
{\mathpzc n}_\pi \ln[ 1+x^2(1-x)^2/\rho^2]\,,
\end{equation}
$\rho = 0.0660$ and ${\mathpzc n}_\pi$ a constant that ensures unit normalisation.  This result is consistent with data \cite{Cui:2021mom} and an array of lattice-QCD calculations \cite{Cui:2022bxn}.  Using the DLY relation, Eq.\,\eqref{EqValenceDF} yields the following elementary FF:
\begin{equation}
\label{EqElemFF}
d_\pi(z;\zeta_{\cal H}) = 0.420 \, z \ln [ 1 + (1-z)^2/(z^4\rho^2)]\,,
\end{equation}
which is drawn in Fig.\,\ref{FReal} (dotted purple curve).  Like that of the original valence quark DF, the dilation of the pion FF is an expression of emergent hadron mass \cite{Roberts:2021nhw, Lu:2022cjx, Ding:2022ows}.
Inserting Eq.\,\eqref{EqElemFF} into the cascade equations, Eq.\,\eqref{SNSFFEqs}, one obtains the numerical solutions that are also drawn in Fig.\,\ref{FReal} (dashed purple curves): qualitatively, they exhibit the same features as the SCI solutions, \emph{viz}.\ those described after Eq.\,\eqref{SolnFFs}.

For use in developing valid insights, the following fits to the numerical solutions can suffice:
%% a1 -> 8.19373, a2 -> -1.83608, a3 -> 4.67305
{\allowdisplaybreaks
\begin{subequations}
\label{EqJetSolutions}
\begin{align}
D_{S_{\mathpzc u}}^\pi(z) & =
%\frac{(1-z) \left[8.19\, -1.84 (1-z)^2\right]}{z[1+4.67 (1-z)] }\,.
%\frac{(1-z)^2 \left(5.59281 z^3-3.56079 z^2+1.52984 z+1.12151\right)}{z \left(0.553066 z^2-1.52801 z+1\right)
\frac{(1-z)^2 \left(1.12 + 1.53 z - 3.56 z^2 + 5.59 z^3\right)}{z \left(1- 1.53 z + 0.553 z^2\right)} \\
D_{N_{\mathpzc u}}^\pi(z) & =
\frac{(1-z)^2 \left(-0.0962+1.42 z +7.45 z^2\right)}{1 - 0.967 z}\,.
%\frac{(1-z)^2 \left(7.44974 z^2+1.42248 z-0.0961816\right)}{1 - 0.967063 z}\,.
\label{EqJetSolutionsA}
\end{align}
\end{subequations}
}

\begin{figure}[t]
\centerline{\includegraphics[width=0.85\columnwidth]{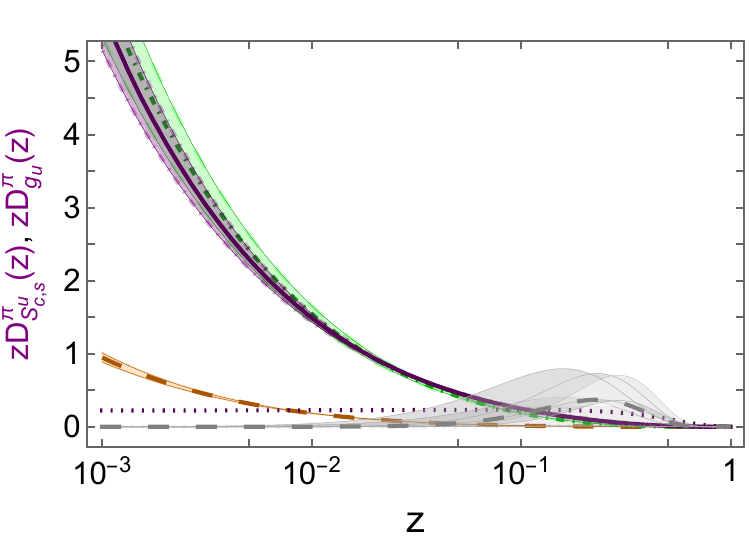}}

\caption{\label{FiggscFFR}
CSM predictions.
Dotted curve -- input $D_{g_{\mathpzc u}}^\pi(z;\zeta_{\cal H})$.
After evolution $\zeta_{\cal H} \to \zeta_1$,
$D_{g_{\mathpzc u}}^\pi(z;\zeta_1)$ -- solid purple curve;
$D_{{S_{\mathpzc s}^{\mathpzc u}}}^\pi(z;\zeta_1)$ -- dot-dashed green;
$D_{{S_{\mathpzc c}^{\mathpzc u}}}^\pi(z;\zeta_1)$ -- long-dashed orange.
For these curves, each bracketing bands obtained by changing $\zeta_{\cal H} \to (1\pm 0.05) \zeta_{\cal H}$.
For comparison, the long-dashed grey curve within like-coloured band shows $D_{g_{\mathpzc u}}^{\pi^+}(z;\zeta_1)$ as obtained from the fit (LO) in Ref.\,\cite{Hirai:2007cx}.
}
\end{figure}

Beginning with the CSM jet FFs drawn in Fig.\,\ref{FReal} (dashed purple curves) and implementing evolution in precisely the manner described in connection with the SCI inputs, including Eq.\,\eqref{GlueAnsatz} with the same value of $\delta=0.113$, then one obtains the forms of $D_{S_{\mathpzc u},N_{\mathpzc u}}(z;\zeta_1)$ drawn in Fig.\,\ref{FReal} (solid purple curves).
(Again, the slight deviation from Eq.\,\eqref{GlueMomFraction} owes to the $s$ and $c$ quark thresholds.)
In these figures, we include an uncertainty on the FFs obtained by changing $\zeta_{\cal H}\to (1\pm 0.05) \zeta_{\cal H}$.
The associated input and evolved gluon FFs are drawn in Fig.\,\ref{FiggscFFR} along with the evolution-generated $s$ and $c$ quark singlet FFs.
Naturally, in all cases, Eq.\,\eqref{FFzone} and its corollaries are satisfied by the CSM FFs.

Here, after starting from the CSM prediction for the elementary FF in Eq.\,\eqref{EqElemFF}:
{\allowdisplaybreaks
\begin{subequations}
\label{MomBeforeAfter}
\begin{align}
\langle z \rangle_{D_{S_{\mathpzc u}}^\pi}^{\zeta_{\cal H}} & = 0.89\,,\;
\langle z \rangle_{D_{g_{\mathpzc u}}^\pi}^{\zeta_{\cal H}} = 0.11\,, \\
%\int_0^1 dz \, z  D_{S_{\mathpzc u}}^\pi(z;\zeta_{\cal H})
%1 & \int_0^1 dz\,z [ D_{g_{\mathpzc u}}^\pi(z;\zeta_{\cal H})]
\langle z \rangle_{D_{S_{\mathpzc u}}^\pi}^{\zeta_{1}} & = 0.78(1)\,,\;
\langle z \rangle_{D_{g_{\mathpzc u}}^\pi}^{\zeta_{1}} = 0.11\,,  \\
\langle z \rangle_{D_{S_{\mathpzc s}^{\mathpzc u}}^\pi}^{\zeta_{1}} & = 0.095(5)\,,\;
\langle z \rangle_{D_{S_{\mathpzc c}^{\mathpzc u}}^\pi}^{\zeta_{1}} = 0.011\,.
\end{align}
\end{subequations}
%where, again, the entries in the last row count, respectively, the momentum fractions in $s+\bar s$ and $c+\bar c$ deriving from the fragmenting $u$ quark.
Despite the vastly different input profiles, highlighting the remarks made in connection with Eq.\,\eqref{MomBalance}, these momentum fractions are unchanged from the SCI values.
}

In Figs.\,\ref{FReal}, \ref{FiggscFFR}, for context, we have again drawn the HKNS fits (LO).  In this case, regarding the light quark FFs, there is qualitative similarity on a large part of the entire domain, especially once low-$z$ fitting uncertainties are taken into account, and fair quantitative agreement on $z \gtrsim 0.4$.  However, concerning the glue FF in Fig.\,\ref{FiggscFFR}, which is poorly constrained by data, little has changed relative to the SCI comparison: the $z$-dependence of the CSM prediction is markedly different from that of the data fit.

%\noindent\emph{5.$\;$Summary}.
%

\section{Summary and Perspective}
\label{Epilogue}
Beginning with an existing prediction for the pion valence quark distribution function (DF), developed using continuum and lattice Schwinger function methods [Eq.\,\eqref{EqValenceDF}];
employing the Drell-Yan-Levy relation [Sec.\,\ref{SecDLY}] to connect this DF with the pion's elementary quark parton fragmentation function (FF) at the hadron scale;
solving a jet hadronisation equation defined therewith [Eq.\,\eqref{SNSFFEqs}];
and subsequently adapting the all-orders evolution $\,$ scheme $\,$ developed for hadron DFs to FFs [Sec.\,\ref{SecFFE}], we delivered parameter-free predictions for pion quark and gluon FFs at a resolving scale $\zeta=1\,$GeV, typically used as a reference in developing fits to data  [Sec.\,\ref{SecFFRealistic}].   In this way, a unified treatment of pion DFs and FFs was accomplished.

Regarding FF evolution, we noted that whilst the evolution equations do not alone ensure momentum conservation for the quark singlet FF, there is a value of the momentum fraction stored in the gluon FF such that,  under evolution, momentum is conserved in the sum over all singlet FFs.  The same value of this momentum fraction ($\approx 11$\% for $4$ quark flavours) achieves momentum conservation for any form of input FFs.

The predicted quark singlet FFs agree qualitatively with existing data fits on the entire domain $z\in (0,1)$, and display quantitative agreement on $z\gtrsim 0.4$.  However, the gluon FFs are markedly different.  In this connection, it should be observed that the paucity of relevant existing data means that gluon FFs are poorly determined in the fits.
More generally, given the large uncertainty in FFs determined via fits to data, the qualitative features of our predictions, including constraints on the large-$z$ behaviour [Eq.\,\eqref{FFzone}], should, at least, provide useful guidance for future such analyses.  This could prove important because improving knowledge of FFs is crucial if best use is to be made of data expected to be gathered at forefront and anticipated facilities.

%In closing, we note that
Extensions of the present analyses to kaon and proton FFs are underway, with a view to developing a unifying set of predictions for hadron FFs that matches in extent those which already exist for hadron DFs \cite{Cui:2020tdf, Lu:2022cjx, Cheng:2023kmt}.
Generalisations to heavy quark FFs are also being explored.

%
%\section*{Acknowledgments}
\medskip
\noindent\emph{Acknowledgments}.
We are grateful to T.~Liu and C.-C.~Ye for valuable discussions.
%% Cong-Cong Ye (叶聪聪) 2020
%O.~Denisov, J.~Friedrich, W.-D.~Nowak and C.~Quintans.
%
%We are grateful for constructive comments from Z.-F.~Cui.
%
Work supported by:
National Natural Science Foundation of China (grant no.\ 12135007);
%
% and
%
Natural Science Foundation of Jiangsu Pro\-vince (grant no.\ BK20220122);
and STRONG-2020 ``The strong interaction at the frontier of knowledge: fundamental research and applications” which received funding from the European Union's Horizon 2020 research and innovation programme (grant agreement no.\ 824093).
%

%\bibliographystyle{../../../zProc/z10/z10KITPC/h-physrev4}
%\bibliographystyle{elsarticle-num-names}
%\bibliography{../../../../CollectedBiB}

\end{document}